\documentclass{llncs}
\usepackage{url}
\usepackage{color}
\usepackage{graphics,graphicx}

\title{Malicious cryptography techniques for unreversable (malicious or not) binaries}
\author{Eric Filiol}
\institute{ESIEA Laval, Laboratoire de cryptologie et de virologie op\'erationnelles, \\
38 rue des Dr Calmette et Gu\'erin, 53000 Laval, France, \\ \url{filiol@esiea.fr}}

\date{}

\begin{document}

\maketitle

\begin{abstract}
Fighting against computer malware require a mandatory step of reverse engineering. As soon as the code has been
disassemblied/de\-com\-pi\-led (including a dynamic analysis step), there is a hope to understand what the malware 
actually does and to implement a detection mean. This also applies to protection of software whenever one wishes to
analyze them. In this paper, we show how to amour code in such a way that reserse engineering techniques (static 
and dymanic) are absolutely impossible by combining malicious cryptography techniques developped in our laboratory 
and new types of programming (k-ary codes). Suitable encryption algorithms combined with new cryptanalytic approaches 
to ease the protection of (malicious or not) binaries, enable to provide both total code armouring and large scale 
polymorphic features at the same time. A simple 400 Kb of executable code enables to produce a binary code and around 
$2^{140}$ mutated forms natively while going far beyond the old concept of decryptor. 
\end{abstract}

%%%%%%%%%%%%%%%%%%%%%%%%%%%%%%%%%%%%%%%
\section{Introduction}
Malicious cryptography and malicious mathematics malicious are an emerging field [2, 3,4] that finds its origin in the 
initial work of Young and Yung on the use of asymmetric cryptography in the design of dedicated offensive functions for 
money extorsion (cryptovirology) [5]. But this initial approach is very limited and gives only a very small insight of 
the way mathematics and cryptography can be actually perverted by malware.

Malicious cryptology and malicious mathematics make in fact explode Young and Yung's narrow vision. This results in an 
unlimited, fascinating yet disturbing field of research and experimentation. This new domain covers several fields and topics
(non-exhaustive list):
\begin{itemize}
\item Use of cryptography and mathematics to develop ``{\em super malware}'' ({\em \"uber-malware}) which evade any kind of detection
      by implementing:
  \begin{itemize}
  \item Optimized propagation and attack techniques (e.g. by using biased or specific random number generator) \cite{roblot}. 
  \item Sophisticated self-protection techniques. The malware code protect itself and its own functional activity by using
        strong cryptography-based tools \cite{filiol_eicar2005}.
  \item Code invisibility techniques using testing simulability \cite{filiol_josse2007}. 
  \item Code mutation techniques (by combining the different methods of the previous three items). 
  \end{itemize}
\item Use of complexity theory or computability theory to design undetectable malware \cite{filiol_hacklu2008}. 
\item Use of malware to perform applied cryptanalysis operations (theft of secrets elements such as passwords or secret keys, 
      static or on-the-fly modification of systems to reduce their cryptographic strength \cite{filiol_dct10} ...) by a direct 
      action on the cryptographic algorithm or its environment. 
\item Design and implementation of encryption systems with hidden mathematical trapdoors. The knowledge of the trap (by the system
      designer only) enables to break the system very efficiently. Despite the fact that system is open and public, the trapdoor must remain
      undetectable. This can also apply to the keys themselves in the case of asymmetric cryptography \cite{Erra_Grenier2009}.
\end{itemize}
To summarize, we can define malicious cryptography and malicious mathematics as the interconnection and interdependence of 
computer virology with cryptology and mathematics for their mutual benefit. The possibilities are virtually infinite. 
It is worth mentioning that this vision and the underlying techniques can be translated for ``non-malicious'' use as the 
protection of legitimate codes (e.g. for copyright protection purposes) against static (reverse engineering) and dynamic
(sandboxing, virtualization) analyses. If simple and classical obfuscation techniques are bound to fail as proved by Barak 
and al. \cite{barak} theoretically (for a rather restricted model of computation), new models of computation \cite{kary,gold}
and new malware techniques \cite{bcps} have proved that efficient code protection can be achieved practically.

If the techniques arising from this new domain are conceptually attractive and potentially powerful, their operational implementation 
is much more difficult and tricky. This requires a very good mastery of algorithms and low-level programming. Above all 
the prior operational thinking of the attack is unavoidable. The same code implementing the same techniques in two different 
contexts will have quite different outcomes. In particular, we must never forget that the code analysis is static 
(disassembly/decompilation) but also dynamic (debugging, functional analysis ...). Code encryption usually protects -- possibly --
against static code analysis only. 

In this paper we will show how the techniques of malicious cryptography enable to implement total amoring of programs, thus prohibiting 
any reverse engineering operation. The main interest of that approach lies in the fact that TRANSEC properties are achieved at the same
time. In other words, the protected binary has the same entropy as any legitimate, unprotected code. This same technique can also 
achieve a certain level of polymorphism/metamorphism at the same time: a suitable 59-bit key stream cipher is sufficient to generate 
up to $2^{140}$ variants very simply. More interestingly, the old fashioned concept of decryptor which usually constitutes a potential 
signature and hence a weakness, is totally revisited.

These techniques have been implemented by the author in the LibThor which has been written and directed by Anthony Desnos 
\cite{libthor1,libthor2}.

The paper is organized as follows. Section~\ref{cacm} recalls some important aspect about code protection and discusses
key points about code armouring and code mutation. Then Section~\ref{cs} presents the context we consider to apply our technique:
we aim at protecting critical portions of code that are first transformed into an intermediate representation (IR), then into the bytecode.
This code itself is that of virtual machine used to provide efficent protection against dynamic analysis. Section~\ref{mprng} presents
the different models and techniques of malicious pseudo-random number generator (PRNG). Finally Section~\ref{imp_k} presents implementation
issue which must be considered to achieve protection against both static and dynamic analyses.
%%%%%%%%%%%%%%%%%%%%%%%%%%%%%%%%%%%%%%%%%%%
\section{Code Armouring and Code Mutation} \label{cacm}
We will not recall in details what these two techniques are. The reader may consult \cite{filiol_bk} for a complete presentation with
respect to them. As far as malicious cryptography is concerned, we are just going to discuss some critical points which are bottlenecks
in their effective implementation. Most of the times the attempts to play with these techniques fail due to a clumsy use of cryptography. 

Code armouring \cite{filiol_eicar2005} consists in writing a code so as to delay, complicate or even prevent its analysis.
As for code mutation techniques they strive to limit (polymorphism) or to remove (metamorphism) any fixed component 
(invariant) from one mutated version to another. The purpose is to circumvent the notion of antiviral signature, may it be a
a simple sequence of contiguous or nor contiguous bytes or meta-structures such as Control Flow Graphs (CFG) and other traces
describing the code functional (behavioral) structure.

In all these cases, the general approach is to encrypt the code to protect or to use special techniques for generating random data. 
But encryption and generation of randomness relates to two major practical problems: their characteristic entropy profile and the secret
elements (keys) management. 
%%%%%%%%%%%%%%%%%%%%%%%%%%%%
\subsection{Entropy profile}
The whole problem lies in the fact that code armouring and code mutation involve random data. These must be generated on-the-fly 
and in the context of metamorphism, the generator itself must be too. For sake of simplicity, we shall speak of {\em Pseudo-Random Number 
Generator} (PRNG) to describe both a random number generator and an encryption system. The difference lies in the fact that in the 
latter case either random data produced from the expansion of the key are combined with the plaintext (stream ciphers) or they are 
the result of the combination of the key with the plaintext (block ciphers). 

The whole issue lies in the generation of a so-called ``good'' randomness. Except that in the context of malicious cryptography \cite{filiol_bk2},
the term ``good'' does not necessarily correspond to what cryptographers usually mean. In fact, it is better -- yet a simplified but sufficient
reduction as a first approximation -- to use the concept of entropy \cite{csw2008}. In the same way, the term of random data will indifferently describe
the random data themselves or the result of encryption.

Consider a (malicious) code as an information source $X$. When parsed, the source outputs characters taking the possible values 
$x_i \quad (i = 0,\ldots , 255)$, each with a probability $p_i = P[X = x_i]$. Then the entropy $H(X)$ of the source is the following
sum\footnote{Let us note that here the entropy considers single characters or 1-grams only. A more accurate value would consider all
the possible $n$-grams and would compute entropy when $n \rightarrow \infty$.}:    
\[ H(X) = \sum_{i = 0}^{255} -p_i\log_2(p_i) \]
Random data, by nature will exhibit a high entropy value\footnote{This means that the uncertainty is maximal whenever
trying to predict the next value output by the source $X$.}. On the contrary, non random data exhibit a low entropy profile 
(they are easier or less difficult to predict). 

From the attacker's point of view\footnote{Do not forget to reverse the view dually: here the attacker is any code 
analyst who wants to have a deep insight into the binary code. However we must keep in mind that what can do a human analyst, 
will be most of the time very difficult (from a computing point of view at least) for automated program as AV software 
are.} the presence of random data means that something is hidden but he has to make the difference between legitimate 
data (e.g. use of packers to protect code against piracy) and illegitimate data (e.g. malware code). In the NATO 
terminology -- at the present time it is the most precise and accurate one as far as InfoSec is concerned-- random 
data relate to a COMSEC ({\em COMmunication SECurity}) aspect only. 

For the attacker (automated software or human expert), the problem is twofold: first detect random data parts inside 
a code and then decrypt them. In this respect, any code area exhibiting a high entropy profile must be considered 
as suspicious. To prevent attention to be focused on those random parts, is it possible to add some TRANSEC ({\em 
TRANSmission SECurity}) aspect. The most famous one is steganography but for malware or program protection purposes it is not directly 
usable (data cannot be directly executed) and we have to find different ways. The other solution is to use malicious 
statistics as defined and exposed in \cite{csw2008}. 

It is also possible to break randomness by using noisy encoding 
techniques like Perseus \cite{perseus}. In this case the code remains executable while being protected AND exhibiting 
low entropy profile at the same time (or the entropy profile of any arbitrary kind of data). As for an exemple, on a 
175-bytes script, the unprotected code has an entropy of 3.90. An encrypted version (by combining simple transpositions 
with simple substitutions) of that code has an entropy equal to 5.5. When protected by Perseus the entropy is about 
2.57. So it is very close to a normal program and thus evade entropy-based detection.

This applies well on any data used for code mutation (e.g. junk code insertion), including specific subsets of code 
as CFGs: randomly mutated CFG must exhibit the same profile as any normal CFG would. Otherwise, considering the COMSEC 
aspect only is bound to make the code detection very easy.
%%%%%%%%%%%%%%%%%%%%%%%%%%%%%
\subsection{Key management} 
Encrypting a code or a piece of code implies its preliminary deciphering whenever it is executed. But in all of the cases -- except 
those involving money extortion introduced Young and Yung \cite{yy} -- the key must be accessible to the code itself 
to decipher. Consequently in a way or another it is contained in a more or less obfuscated form inside the code. Therefore 
is it accessible to the analyst who will always succeed in finding and accessing it. Instead of performing cryptanalysis, 
a simple decoding/deciphering operation is sufficient. 
 
It is therefore necessary to consider keys that are external to the encrypted code. Two cases are possible \cite{filiol_bk2}:
\begin{itemize}
\item Environmental key management introduced in \cite{riordan} and developped in \cite{filiol_bk2}. The code gathers information 
      in its execution environment and calculates the key repeatedly. The correct key will be computed when and only when the suitable
      conditions will be realized in the code environment -- which is usually under the control of the code designer. The security 
      model should prohibit dictionary attacks or environment reduction attacks (enabling reduced exhaustive search) by the code 
      analyst. Consequently the analyst must examine the code in an controlled dynamic area (sandbox or virtual machine) and wait 
      until suitable conditions are met without knowing when they will be. However it is possible to build more operational scenarii
      for this case and to detect that the code is being analyzed and controlled \cite{filiol_bk2}.
\item Use of $k$-ary codes \cite{karyd,kary} in which a program is no longer a single monolithic binary entity but a set of binaries
      and non executable files (working in a serial mode or in a parallel mode) to produce a desired final (malicious or not) action. 
      Then the analyst has a reduced view on the whole code only since generally he can access a limited subset of this $k$-set. 
      In the context of (legitimate) code protection, one of the files will be a kernel-land module communicating with a userland 
      code to protect. The code without the appropriate operating environment -- with a user-specific configuration by the 
      administrator -- will never work. This solution has the great advantage of hiding (by outsourcing it), the encryption 
      system itself. It is one particular instance with respect to this last solution that we present in this paper.
\end{itemize}
%%%%%%%%%%%%%%%%%%%%%%%%%%%%%%%
\section{Case Studies} \label{cs}
In this section we are going to illustrate our approach with simple but powerful cases. Without loss of generality and 
for sake of clarity, we consider a reduced case in which only a very few instructions are protected against any
disassembly attempt. Of course any real case -- as we did in LibThor -- will consider far more instructions, especially
all critical ones (those defining the CFG for instance).
%%%%%%%%%%%%%%%%%%%%%%%%%%%%%%%%%%
\subsection{The working context}
The LibThor library initiated and mainly developped by Anthony Desnos \cite{libthor1,libthor2} uses virtual
machines (VM) as the core tool. Virtual Machines offer a powerful protection for an algorithm or anything
else that we would like to protect against reverse engineering. We can have a lot of different VMs piled up 
(like Russian dolls) into 
a software. This technique must be combined with classical techniques since it is just a one more step 
towards code protection.

In LibThor, the main goal of VMs is to build a simple code which interprets another one. The idea is to 
take a piece of x86 assembly instructions and to have a simple, dynamic, metamorphic VMs very quickly to interpret
it. It is worth stressing on the fact that you can embed different VM into the same program to protect differents parts.

In Desnos' LibThor VMs are codes which interpret an intermediate representation (IR) derived from the REIL 
language \cite{reil}. So a translation between x86 assembly instructions and the chosen IR is performed. 
The VM is totaly independent 
from the LIBC or anything like that. The main goal of the VM is to run an algorithm into an encapsulated
object (which can be loaded anywhere; in fact it is a simple shellcode), but the main feature of the VM 
is to have a different version of it at each generation for a same code so if we want to protect a simple
instruction $X$, we can build a new VM every time to protect the same $X$. This new VM will be different from the previous 
one.

Moreover everything must be dynamic. Therefore we must have simple working rules:
\begin{itemize}
\item the bytecode must be dynamic and hidden by using the malicious cryptography tools (malicious PNRG) we are exposing
      hereafter;
\item information must be dynamic (area for each context, the offset for each register in the context...);
\item the code must be transformed with the internal LibThor metamorphic package (in which malicious PNRG can used used too);
\item integer constants can be hidden with LibThor internal junk package. This is very useful to hide opcode values, 
      register offsets or anything which represent any invariant in a program. Here again malicious PRNGs play a critical 
      role.
\end{itemize}

During the code generation (mutation), we use the LibThor metamorphic package a lot of times on a function.
We have several steps during a generation:
\begin{enumerate}
\item Obfuscation of C source code.
\item Compiling the C source code of the VM into a library.
\item Extraction of interesting functions from the library.
\item Transformation of x86 assembly instructions into IR.
\item Obfuscation of IR by using metamorphic package.
\item Transformation of IR into bytecode.
\item Creation of dynamic functions to handle the bytecode.
\item Obfuscation of functions by using the metamorphic package. The mutated code is produced here and the 
      malicious PRNG is mainly involved here.
\item Assembly of all parts to have a simple shellcode.
\end{enumerate}
Our PRNG is essentially dedicated to the protection of the bytecode which is the final result of the
transformation: {\tt X86 ASM} $\rightarrow$ {\tt REIL IR} $\rightarrow$ {\tt bytecode}. Here is an 
illustrating example: 
\begin{verbatim}
   [X86 ASM]       MOV EAX, 0x3 [B803000000]
   [REIL IR]       STR (0x3, B4, 1, 0), (EAX, B4, 0, 0)
   [BYTECODES]     0xF1010000 0x40004 0x3 0x0 0x6A
\end{verbatim}

We have five fields in the bytecode corresponding respectively to :
\begin{itemize}
\item  {\tt 0xF1010000}
  \begin{itemize}
  \item {\tt 0xF1}: the opcode of the instruction (STR),
  \item {\tt 0x01}: specifies that it is an integer value,
  \item {\tt 0x00}: useless with respect to this instruction,
  \item {\tt 0x00}: specifies that it is a register.
  \end{itemize}
\item {\tt 0x40004}
  \begin{itemize}
  \item {\tt 0x04}: the size of the first operand,
  \item {\tt 0x00}: useless with respect this instruction,
  \item {\tt 0x04}: the size of the third operand,
  \end{itemize}
\item {\tt 0x3}: direct value of the integer,
\item {\tt 0x0}: useless with respect to this instruction,
\item {\tt 0x6A}: value of the register.
\end{itemize}
In this setting the {\tt 0x00} values (useless with respect this instruction) contribute directly to the TRANSEC aspect.
Now that the working environment is defined, let us explain how a malicious PRNG can efficiently protect
the opcode values while generating them dynamically.
%%%%%%%%%%%%%%%%%%%%%%%%%%%%%%%
\section{Malicious PRNG} \label{mprng}
Sophisticated polymorphic/metamorphic or obfuscation techniques must rely on 
PRNG (Pseudo-Random Number Generator). In the context of this paper, the 
aim is to generate sequences of random numbers (here bytecode values) on-the-fly
while hiding the code behavior. 

Sequences are precomputed and we have to design a generator which will 
afterwards output those data. The idea is that any data produced by the 
resulting generator will be first used by the code as a valid address, and
then will itself seed the PNRG to produce the next random data.

Three cases are to be considered:
\begin{enumerate}
\item the code is built from any arbitrary random sequence;
\item the sequence is given by a (non yet protected) instance 
      of bytecode and we have to design an instance of PNRG accordingly;
\item a more interesting problem lies in producing random data that can be
      somehow interpreted by a PRNG as meaningful instructions like 
      {\tt jump 0x89} directly.
\end{enumerate} 
This relates to interesting problems of PRNG cryptanalysis. We are going 
to address these three cases.

From a general point of view it is necessary to recall that for both three
cases the malware author needs reproducible random sequences. By reproducible
(hence the term of pseudo-random), 
we mean that the malware will replay this sequence to operate
its course of actions. The reproducibility condition implies to consider a 
{\em deterministic Finite-State Machine} (dFSM). The general scheme of how this 
dFSM is working is illustrated as follows. Without the dFSM, any instruction 
data whenever executed produced a data used the next instruction (e.g. an 
address, an operand...). 
\[ I_0 \rightarrow D_0 \rightarrow I_1 \rightarrow D_1 \ldots \rightarrow D_i \rightarrow I_(i+1) \rightarrow  \ldots \]

Any analysis of the code will easily reveal to the malware analyst all the
malware internals since all instructions are hardcoded. But if a few 
data/instructions are kept under an encrypted form, and deciphered at 
execution only, the analysis is likely to be far more difficult (up to decryptor and the secret
key protection issue). It is 
denied of a priori analysis capabilities. So we intend to have
\[ I_0 \rightarrow D'_0 \rightarrow I_1 \rightarrow D_1 \ldots \rightarrow D'_i \rightarrow I_(i+1) \rightarrow  \ldots \]
where $dFSM(D'_i) = D_i$ for all i. Upon execution, we just have to input 
data $D'_i$ into the dFSM which will then output the data $D_i$. A few critical points are 
worth stressing on
\begin{enumerate}
\item no key is neither required nor used; 
\item instructions can similarly protected as well.
\end{enumerate}
Of course to be useful as a prevention tool against analysis, the dFSM 
must itself be obfuscated and protected against analysis. But this last 
point is supposed to be fulfilled. We will show a more powerful implementation
by using $K$-ary malware in Section~\ref{imp_k}.
%%%%%%%%%%%%%%%%%%%%%%%%%%%%%%%%%%%%%%%%%%%%%%%%%%%%%
\subsection{(Malware) Code Built From an Arbitrary Sequence}
In this case, the sequence is arbitrary chosen before the design of the 
code and hence the code is written directly from this arbitrary sequence. 
This case is the most simple to manage. We just have to choose carefully 
the dFSM we need. One of  the best choice is to take a congruential 
generator since it implies a very reduced algorithm with simple 
instructions.

Let us consider $X_0$ an initial value and the corresponding equation
\[ x_(i+1) = a*X_i + b \qquad mod(N) \]
where $a$ is the multiplier, $b$ is the increment and $N$ is the modulus. 
Since the length of the sequence involved in the malware design is rather very 
short (up to a few tens of bytes), the choice of those parameters is not 
as critical as it would be for practical cryptographic applications.
In this respect, one can refer to Knuth's seminal book to get the 
best sets of parameters \cite{knuth}.

Here are a few such examples among many others:
\begin{description}
\item[Standard minimal generator] $a = 16,807 - b = 0 - N = 2^{31} - 1$. 
\item[VAX-Marsaglia generator] $a = 16,645 - b = 0 - N = 2^{32}$. 
\item[Lavaux \& Jenssens generator] $a = 31,167,285 - b = 0 - N = 2^{48}$.
\item[Haynes generator] $a = 6,364,136,223,846,793,005 - b = 0 - N = 2^{64}$.
\item[Kuth's generator] $a = 22~695~477 - b = 1 - N = 2^{32}$ and $X_{n + 1} >>= 16$.
\end{description}
Of course the choice of the modulus is directly depending on the data type used in the malware.

Another interesting approach is to consider hash functions and S/key. The 
principle is almost the same. We take a $(m,n)$ hash function $H$ which 
produces a $n$-bit output from a $m$-bit input with $m > n$. In our case 
we can build $m$ in the following way
\begin{verbatim}
 m = <data to protect><padding of random data><size of data>
\end{verbatim}
or equivalently
\begin{verbatim} 
 m = D_i <random data> |D_i|
\end{verbatim}
Then we choose a $m$-bit initialization vector (IV) and we compute the random 
sequence as follows
\[ IV \rightarrow D_i = H(IV) \rightarrow x = H^{|D_i|}(D_i) \rightarrow y = H^{|x|}(x) \rightarrow H^{|y|}(y) \rightarrow \]
The iteration value $|D_i|$ can be used to get one or more required arbitrary 
value thus anticipating the next case. Of course the nature of the hash 
function is also a key parameter: you can either use existing hash function 
(e.g MD5, SHA-1, RIPEMD 160, SHA-2...) and keep only a subset of the output 
bit; or you can design your own hash function as explained in \cite{knuth}.
%%%%%%%%%%%%%%%%%%%%%%%%%%%%%%%%%%%%%%%%%%%%%%%%%%%%%
\subsection{Random Sequence Coming From a Arbitrary (Malware) Code}
In this slightly different case, the sequence is determined by a (non yet protected) instance
of a code. This issue is then to design or use an instance of PRNG accordingly. 
This is of course a far more difficult issue which implies cryptanalytic 
techniques. To formalize the problem we have a sequence 
\[ X_0, X_1, X_2 \ldots x_i \ldots X_n \]
which represents critical data (addresses, ASM instructions, operands...) 
of a particular instance of a (malware) code. As for example let us consider 
three series of 32-bit integers describing bytecode values as defined in Section~\ref{cs}:
\begin{verbatim}
0x2F010000 0x040004 0x3 0x0 0x89        (1)
0x3D010000 0x040004 0x3 0x0 0x50        (2)    
 0x5010000 0x040004 0x3 0x0 0x8D        (3)  
\end{verbatim}
They are just different instances of the same instruction \cite{libthor2}.
The aim is to have these data in the code under a non hardcoded form. Then 
we intend to code them under an obfuscated form, e.g.
\[ K_0, K_1, K_2, \ldots K_i, \ldots K_n \ldots \]
We then have to find a dFSM such that 
\[ X_0 = dFSM(K_0), X_1 = dFSM(K_1) \ldots X_i = dFSM(K_i) \ldots \]
The notation $K_i$ directly suggests that the quantity input to the dFSM is a
key in a cryptographic context but these keys have to exhibit local low entropy profile at the same
time. So the malicious PRNG must take this into account as well. In this case, we have to face a two-fold cryptanalytic 
issue:
\begin{itemize}
\item either fix the output value $X_i$ and find out the key $K_i$ which outputs 
      $X_i$ for an arbitrary dFSM, 
\item or for an arbitrary set of pairs $(X_i, K_i)$ design a unique suitable 
      dFSM for those pairs.
\end{itemize}
The first case directly relates to a cryptanalytic problem while the second
refers more to the problem of designing cryptographic dFSMs with trapdoors.
In our context of malicious cryptography, the trapdoors here are precisely 
the arbitrary pairs of values $(X_i, K_i)$ while the dFSM behaves for any 
other pair as a strong cryptosystem \cite{eciw2010}. This second issue is 
far more complex to address and will not be exposed in this paper (research 
under way; to be published later). 

We will focus on the first case which has been partially addressed for 
real-life cryptosystem like {\em Bluetooth} E0 \cite{e0} in the context 
of zero knowledge-like proof of cryptanalysis. But in our case we do not 
need to consider such systems and much simpler dFSM can be built 
conveniently for our purposes: sequences of data we use are rather short.
       
We have designed several such dFSMs and we have proceeded to their 
cryptanalysis to find out the keys $K_i$ which output the values $X_i$. 

As we will see, those dFSM have to exhibit additional features in order to
\begin{itemize}
\item be used for code mutation purposes,
\item exhibit TRANSEC properties. In other words, if we have $Y = dFSM(X)$, then $X$ and $Y$
      must have the same entropy profile. Replacing $X$ with a $Y$ having a higher entropy profile
      would focus the analyst's attention. 
\end{itemize}
Without loss of generality, let us consider the mathematical description of a 59-key bit stream cipher 
(we currently work on block cipher which should offer more interesting 
features; to be continued \ldots) we have designed for that purpose. Other dFSMs with larger key size
(up to 256) have been also designed (available upon request). 

\begin{figure}[htbp]
\centering
\includegraphics[width=\textwidth]{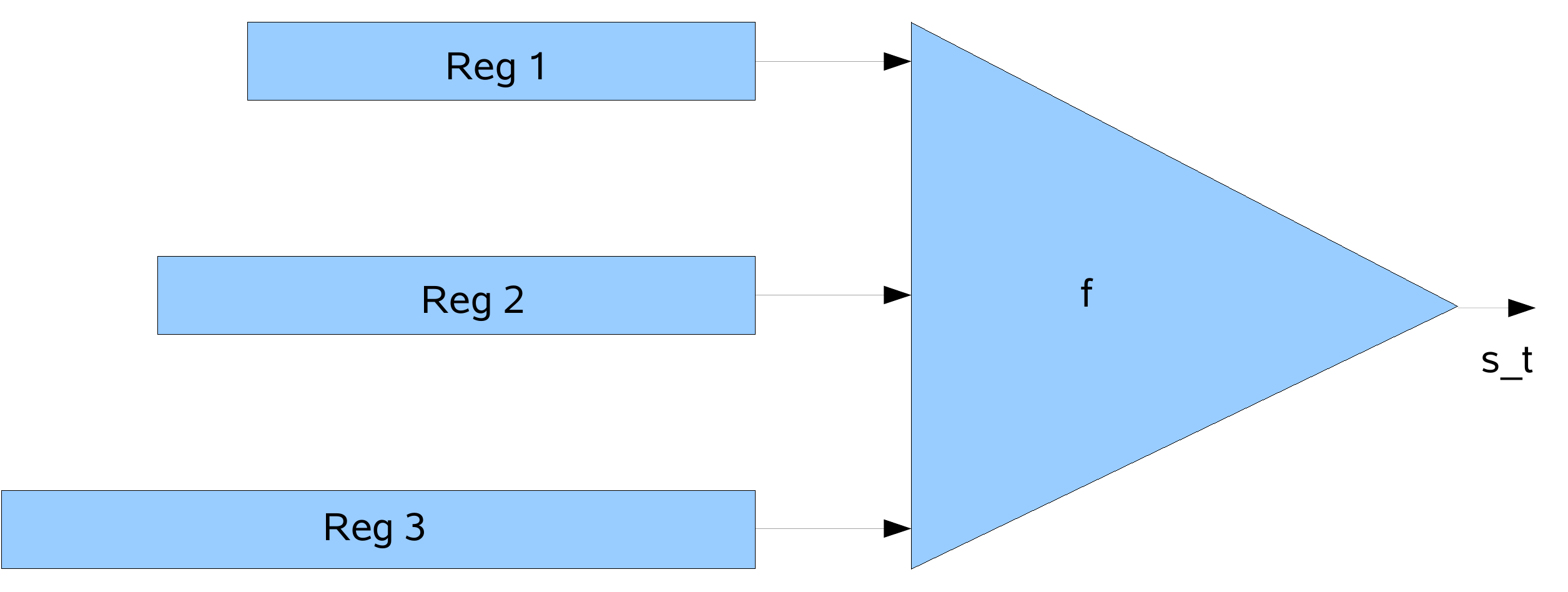}
\caption{Malicious 59-bit key deterministic finite state-machine (stream cipher)} \label{combsys}
\end{figure}

The principle is very simple yet powerful: three registers $R_1, R_2$ and $R_3$ are
filtered by a Boolean function $F$, thus outputing bits $s_t$ (or bytes in a
vectorized version) that are combined to the plaintext (Figure~\ref{combsys}).
The value $K_i$ initializes the content of registers $R_1, R_2$ and $R_3$ 
at time instant $t = 0$, and outputs bits $s^t$ which represent the 
binary version of values $X_i$.

The linear feedback polynomials driving the registers are the following:
\begin{eqnarray*}
P_1(x) & = & x^{17} \oplus x^{15} \oplus x^{14} \oplus x^{13} \oplus x^{11} \oplus x^{10} \oplus x^{9} \oplus x^{8} 
             \oplus x^6 \oplus x^5 \oplus x^4 \oplus x^2 \oplus 1                             \\
P_2(x) & = & x^{19} \oplus x^{18} \oplus x^{16} \oplus x^{15} \oplus x^{11} \oplus x^{10} \oplus x^5 \oplus x^3 \oplus x^2 \oplus x \oplus 1 \\    
P_3(x) & = & x^{23} \oplus x^{22} \oplus x^{21} \oplus x^{20} \oplus x^{17} \oplus x^{16} \oplus x^{15} \oplus x^{12} \oplus x^{10} \oplus x^{8} \oplus x^7 \oplus x \oplus 1
\end{eqnarray*}
where $\oplus$ denotes the exclusive OR.

The combination function is the {\em Majority function} applied on three input bits and is given by
\[ f(x_1, x_2, x_3) = x_1x_2 \oplus x_1x_3 \oplus x_2x_3. \]
An efficient implementation in C language is freely available by contacting the author. 
For our purposes, we will use it in a procedure whose prototype is given by
\begin{verbatim}
void sco(unsigned long long int * X, unsigned long long int K)
 {
  /* K obfuscated value (input), X unobfuscated value (output) */
  /* (array of 8 unsigned char) by SCO                         */
  ...
 }
\end{verbatim}
Now according to the level of obfuscation you need, different ways exist
to protect your critical data inside the code (series of integers (1), (2) 
and (3) above). We are going to detail two of them.
%%%%%%%%%%%%%%%%%%%%%%%%%%%%%%%%%%%%%%%%%%%%%%%%%%%%
\subsubsection{Concatenated bytecodes}
The dFSM outputs critical data under a concatenated form to produce chunks 
of code corresponding to the exact entropy of the input value $(K_i)$, thus 
preventing any local increase of the code entropy. For the dFSM considered, 
it means that we output series (1), (2) and (3) under the following form 
\begin{verbatim}
  1)--> 0x2F01000000040004000000030000000000000089      
  2)--> 0x3D01000000040004000000030000000000000050
  3)--> 0x050100000004000400000003000000000000008D  
\end{verbatim}
Let us detail the first output sequence (1). It will be encoded as three 
59-bit outputs $M_1, M_2$ and $M_3$
\begin{verbatim}     
   M_1 =     0x0BC04000000LL;
   M_2 = 0x080008000000060LL;
   M_3 = 0x000000000000089LL; 
\end{verbatim}
To transform $M_1, M_2$ and $M_3$ back into five 32-bit values $X_1, X_2, 
X_3, X_4$ and $X_5$, we use the following piece of code:
\begin{verbatim}      
  /* Generate the M_i values */
  sco(&M_1, K_1);
  sco(&M_2, K_2);
  sco(&M_3, K_3);

  X_1 = M_1 >> 10;  /* X_1 = 0x2F010000L */       
  X_2 = ((M_2 >> 37) | (M_1 << 22)) & 0xFFFFFFFFL 
                    /* X_2 = 0x00040004L */
  X_3 = (M_2 >> 5) & 0xFFFFFFFFL; /* X_3 = 0x3 */
  X_4 = ((M_3 >> 32) | (M_2 << 27)) & 0xFFFFFFFFL;
                    /* X_4 = 0x0 */
  X_5 = M_3 & 0xFFFFFFFFL;       /* X_5 = 0x89 */
\end{verbatim}      
Values $M_1, M_2$ and $M_3$ will be stored in the code as the values
$K_1, K_2$ and $K_3$ with $dFSM(K_i) = M_i$:
\begin{verbatim}
   K_1 = 0x6AA006000000099LL;
   K_2 = 0x500403000015DC8LL;
   K_3 = 0x0E045100001EB8ALL;    
\end{verbatim}
Similarly we have for sequence (2) 
\begin{verbatim}         
   M_1 =     0x0F404000000LL;  K_1 = 0x7514360000053C0LL;
   M_2 = 0x080008000000060LL;  K_2 = 0x4C07A200000A414LL;
   M_3 = 0x000000000000050LL;  K_3 = 0x60409500001884ALL;
\end{verbatim}
and for sequence (3)
\begin{verbatim}
   M_1 =     0x01404000000LL;  K_1 = 0x76050E00001F0B1LL;
   M_2 = 0x080008000000060LL;  K_2 = 0x00000010C80C460LL;
   M_3 = 0x00000000000008DLL;  K_3 = 0x000000075098031LL;
\end{verbatim}
The main interest of that method is that the interpretation of code  
is not straightforward. Code/data alignment does not follow any logic 
(that is precisely why we have chosen a 59-bit FSM which is far better 
that a seemingly more obvious 64-bit FSM ; any prime value is optimal). 

Moreover, as we can notice, the $K_i$ values are themselves sparse as unobfuscated
opcodes are (structural aspect). Additionally, their entropy profile (quantitative aspects) 
is very similar to the $M_i$ values (and hence the $X_i$ ones). This implies that any 
detection techniques based on local entropy picks is bound to fail.
      
Due to the careful design of our 59-bit dFSM, we managed to obtain a 
unicity distance which is greater than 59 bits (the unicity distance 
is the minimal size for a dFSM output to be produced by a single secret 
key). In our case a large number of different 59-bit keys can output 
an arbitrary output sequence. Here are the results for the three series 
(1), (2) and (3) (Table~\ref{res}):
\begin{table}[htbp]
\begin{center}
\begin{tabular}{c|c|c} \hline
 Serie   &  $M_i$ values     &  Number of secret keys $K_i$    \\ \hline \hline
  (1)    &   $M_1$           &      314    (file res11)        \\
  (1)    &   $M_2$           &     2,755   (file res12)        \\
  (1)    &   $M_3$           &     8,177   (file res13)        \\ \hline
  (2)    &   $M_1$           &      319    (file res21)        \\
  (2)    &   $M_2$           &     2,755   (file res22)        \\
  (2)    &   $M_3$           &    26,511   (file res23)        \\ \hline
  (3)    &   $M_1$           &     9,863   (file res31)        \\
  (3)    &   $M_2$           &     2,755   (file res32)        \\
  (3)    &   $M_3$           &     3,009   (file res33)        \\ \hline
\end{tabular}
\caption{Number of possible keys for a given output value} \label{res}
\end{center}
\end{table}

This implies that we can randomly select our 9 $M_i$ values and thus 
we have 
\begin{eqnarray*}
 &   & 314 \times (2,755)^3 \times 8,177 \times 319 \times 26,511 \times 9,863 \times 3,009  \\
 & = & 13,475,238,762,538,894,122,655,502,879,250 
\end{eqnarray*}
different possible code variants. It is approximatively equal to $2^{103}$ 
variants. The different files ($resij$ with $i, j \in \{1, 2, 3\}$) 
whose name is given in right column of Table~\ref{res} are freely 
available upon request.

To build a variant, you just have to select data in the nine files randomly 
according to the following piece of code (code extract to generate the $M_1$ 
values for the first serie (1) only; refer to Section~\ref{imp_k} for a 
secure implementation):
\begin{verbatim}
  f1 = fopen("res11","r");         
  f2 = fopen("res12","r");
  f3 = fopen("res13","r");
                                 
  randval = (314.0*(rand()/(1 + RAND_MAX));     
  for(i = 0; i < randval; i++) 
     fscanf(f1, “%lX %lx %lx\n”, &y1,&y2, &y3);            
  K_1 = y1 | (y2 << 17) | (y3 << 36) ;         
   /* do the same for values M_2 and M_3 of serie (1) */
      ....             
   /* repeat the same for series (2) and (3)          */
      ....             
   /* Generate M_1 value for series(1)                */             
   sco(&M_1, K_1);
\end{verbatim}
The different files ($resij$ with $i, j \in \{1, 2, 3\}$) can be stored
in a program whose size is less than 400 Kb.
%%%%%%%%%%%%%%%%%%%%%%%%%%%%%%%%%%%%%%%%%%%%%%%%%%%%
\subsubsection{Non concatenated bytecodes}
In this second case, the dFSM outputs 59-bit chunks of data whose 
only the 32 least significant bits are useful. In this case we 
output five 59-bit chunks of data $M_1, M_2, M_3, M_4$ and $M_5$. 
For sequence (1) we have
\begin{verbatim}   
   M_1 = 0x???????2F010000LL; 
   M_2 = 0x???????00040004LL; 
   M_3 = 0x???????00000003LL; 
   M_4 = 0x???????00000000LL; 
   M_5 = 0x???????00000089LL;
\end{verbatim}
where \verb+ ?+ describes any random nibble. We get the $X_i$ values 
from the $M_i$ values with $X_i = M_i \quad \& \quad 0xFFFFFFFFL;$ 

The main interest of that method is that it naturally and very simply 
provides increased polymorphism compared to the previous approach.
Indeed about $2^{140}$ 5-tuples $(K_1, K_2, K_3, K_4, K_5)$ whenever input
in our dFSM produces 5-tuples $(X_1, X_2, X_3, X_4, X_5)$. Then we can
produce a huge number of different instances of the same code by
randomly choosing any possible 5-tuples. By increasing size of the
memory of the FSM we even can arbitrarily increase the number of 
possible polymorphic instances.
%%%%%%%%%%%%%%%%%%%%%%%%%%%%%%%%%%%%%%%%%%%%%%%%%%%%%%%%%%%%%%%%%%%%%%%%%%%%%%
\section{Operational Implementation} \label{imp_k}
The obfuscation techniques we have presented before, which are based on malicous
cryptography (malicious, biased PRNG) and cryptanalysis techniques (to precompute
inputs to the dFSM) require obviously to protect the dFSM itself very strongly. 
Indeed obtaining the $X_i$ values from the $K_i$ is straightforward whenever we 
have the dFSM. We would then come back to the weak existing implementations and
reduce the problem to a simple decoding scheme.

But this obfuscation is impossible to reverse in the context of $k$-ary malware 
\cite{karyd,kary}. In this case, the viral information is not contained in a 
single code as usual malware do, but it is split into $k$ different innocent-looking 
files whose combined action -– serially or in parallel -– results in the actual 
malware behavior and in the metamorphic code instance generation.

\begin{figure}[htbp]
\centering
\includegraphics[width=\textwidth]{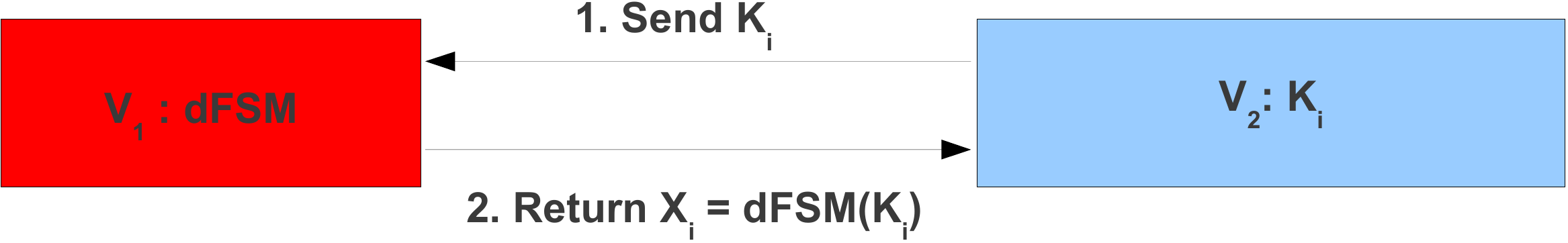}
\caption{2-ary implementation of malicious PRNG} \label{fig2}
\end{figure}

Just imagine a 2-ary code (we can take of course $k > 2$) made of parts $V_1$
and $V_2$. Part $V_1$ just embeds the dFSM and wait in memory for values $K_i$ 
coming from part $V_2$. Its role is to compute $dFSM(K_i)$ values and send 
them back to part $V_2$ according to the protocol summarized by Figure~\ref{fig2}.
The interpretation of data by the reverser (and hence the reversing of the code) 
is no longer possible since the dFSM is deported in the part $V_1$ and the analysis 
of part $V_2$ would require to reconstruct the dFSM and to break it. This is of 
course impossible since the code does not contain a sufficient quantity of
encrypted information.

In the case of metamorphism, the part $V_1$ will output a random $K_i$ value instead during
the code instance generation.

From a practical point of view, we have considered different implementations. 
\begin{description}
\item[Communication pipes] Due to the relationship between father and children processes, only parallel class A or C
      \cite{kary} k-ary codes can be implemented. It is not the most optimal solution.
\item[Named communication pipes] In this case, k-ary parallel class B codes can be efficiently implemented (the most 
     powerful class since there is no reference in any part [hence information] to other any part). 
\item[System V IPC] This is the most powerful method since everything is located into shared memory.
\end{description}
The source codes of our implementation will be soon publicly available.

In the context of code protection for legitimate purpose, the part $V_2$ will be a kernel-land
module can be user/host specific. It can also be a program located on a server outside the code
analyst's scope.
%%%%%%%%%%%%%%%%%%%%%%%%%%%%%%%%%%%%%%%%%%%%%%%%
\section{Conclusion}
We have shown in this paper that it is possible to prevent code analysis from reversing while ensuring a high level
of metamorphism. Of course, we have to apply those techniques to all critical parts of the code (especially those
which determine the execution flow). In this case it is obvious that static analysis is no longer possible. As
far as dynamic analysis is concerned, the analyst has to perform debugging and sandboxing. But using delay detection
technique \cite{filiol_bk2} can trigger a different code behaviour and code mutation.

Our current research and experimentation work is related to non deterministic FSMs. In this case any internal state of 
the FSM results in many possible outcome (next state at time instant $t+1$). Even if is likely to be far beyond any AV software
detection capability, human analysis becomes impossible. Non deterministic contexts result in untractable complexity.
%%%%%%%%%%%%%%%%%%%%%%%%%%%%%%%%%%%%%%%%%%%%%%%%
\section*{Acknowledgement}
Thanks to Anthony Desnos for his support with respect to virtual machines and IR and for many other reasons: Geoffroy Gueguen
(and his awful vanWinjgaarden formal grammars), Eloi Vanderb\'eken (aka {\em Baboon}) for his funny way of thinking security.
%%%%%%%%%%%%%%%%%%%%%%%%%%%%%%%%%%%%%%%%%%%%%%%%

\end{document}